\documentclass[manuscript, authorversion]{acmart}

\usepackage{multirow}
\usepackage{graphicx}
\usepackage[table,xcdraw]{xcolor}
\usepackage{subcaption}

\AtBeginDocument{%
  }

\begin{document}

\author{A'di Dust}
\author{Mohammad Soleymani}
\author{Mina Kian}
\author{Maja Matari\'c}

\title[Synchrony, Affect, and Perceived Autonomy]{Synchrony, Affect, and Perceived Autonomy \\in Virtual Robot-Facilitated Anxiety Support Groups}

\begin{abstract}
Socially assistive robots could expand access to group-based mental health support when skilled human facilitators are unavailable. However, their effectiveness will depend on whether they foster productive interactions among participants. We examined dyadic head-movement synchrony among 96 participants across 18 virtual robot-facilitated anxiety support group sessions. A human-controlled virtual robot, via a Wizard-of-Oz (WoZ) system, facilitated all sessions. We found that stronger peak synchrony predicted lower post-session negative affect. In contrast, speakers who more often followed listeners’ movements reported higher negative affect and state anxiety, demonstrating that synchrony strength and direction capture distinct interpersonal processes. Participants who believed the robot was autonomous showed more genuine synchrony and stronger peaks than those who believed it was human-operated; more negative attitudes toward robots also predicted greater synchrony. These findings support synchrony as an objective, behaviorally grounded measure for evaluating robot-facilitated groups and a potential signal for guiding future autonomous facilitators.
\end{abstract}

\ccsdesc[500]{Human-centered computing~Empirical studies in HCI}

\keywords{HRI, Socially Assistive Robots, Dyadic Interaction, Synchrony, Anxiety, Support Groups.}

\maketitle

\section{Introduction}
Motivated by the high rates of anxiety in the population, there is a growing trend of developing AI and robots for use in mental health contexts \cite{cruz2025artificial}.
However, successful mental health interventions--individual or group--require engaged interactions among participants \cite{davison2000talks, burlingame2011cohesion}, and it remains unclear whether such interpersonal processes emerge when artificial agents facilitate these interactions.

Past work on AI- and robot-facilitated mental health interventions has typically focused on traditional therapy. Individual and group therapy involves the structured delivery and discussion of clinical intervention tools by licensed clinicians, posing challenges for non-human agents. To address this challenge, we focus on anxiety support groups, which are not clinical interventions but instead center on shared experiences, peer interaction, and mutual assistance \cite{beard2024evaluation}. Despite this distinction from traditional therapy, support groups and group therapy rely on related interpersonal processes, and substantially more research has examined the facilitation of group therapy than support groups.

When designing AI or robotic agents to facilitate support groups, it is important to consider both facilitator-participant and participant-participant interactions. Although the therapeutic alliance between facilitators and participants, especially in dyadic therapy, has been studied extensively \cite{ardito2011therapeutic}, much less research has examined interactions among participants, particularly in support groups. Technologies designed for these settings must therefore identify and support positive participant interactions, not merely deliver therapeutic content.

One relevant measure of participant interaction is {\it nonverbal synchrony}--the temporal, often spontaneous coordination of movement among participants \cite{palumbo2017interpersonal}. Head movement is an established measure of nonverbal synchrony \cite{ramseyer2014nonverbal}, motivating its use in this work. We measured synchrony using windowed cross-correlation with peak picking and generated surrogate time series to screen for pseudosynchrony \cite{moulder2018determining}. 
Synchrony is especially useful because it provides an \textit{observable} marker of meaningful interpersonal processes rather than relying solely on self-report or inherently unobservable constructs. Beyond its value as an evaluation metric, synchrony could serve as an observable objective function for future autonomous facilitators, enabling them to adapt their behavior to support productive participant interaction.

Existing evidence links behavioral and nonverbal synchrony to outcomes in human-facilitated therapy \cite{atzil2023facilitating, ramseyer2011nonverbal, paulick2018nonverbal, altmann2020associations}. However, it remains unclear whether this relationship extends to robot-facilitated contexts, especially in group settings. Finding a similar relationship in robot-facilitated mental health support would suggest that clinically relevant interaction processes can emerge around artificial facilitators, supporting further research on robot- and AI-facilitated support groups. This motivates our first research question:

\begin{itemize}
    \item \textbf{RQ1}: Is the relationship between head-movement synchrony and therapeutic outcomes in simulated robot-facilitated anxiety support groups consistent with relationships previously reported in human-facilitated groups?
\end{itemize}

A second consideration in designing robot- and agent-facilitated support groups is whether a robot facilitator changes participants’ behavior compared to a human facilitator. Participants’ behavior may depend on how they understand the robot’s autonomy, as well as on their preexisting attitudes toward robots. This motivates our second research question:

\begin{itemize}
    \item \textbf{RQ2}: Are participants’ perceptions of the simulated robot associated with differences in nonverbal synchrony?
\end{itemize}

We investigated whether simulated robot facilitators preserve interaction qualities associated with effective human-facilitated support groups. We analyzed dyadic head-movement synchrony among 96 participants across 18 simulated robot-facilitated anxiety support-group sessions. All sessions used a Wizard-of-Oz (WoZ) system and experimentally manipulated participants’ beliefs about the robot’s autonomy. In nine sessions, participants were told that the robot was remotely operated; in the other nine, they were told that it operated autonomously. We examined how dyadic synchrony related to positive affect, negative affect, and state anxiety, and whether perceived robot autonomy and negative attitudes towards robots influenced interaction dynamics.

Our findings revealed patterns consistent with prior research on human-facilitated psychotherapy \cite{atzil2023facilitating, ramseyer2011nonverbal, paulick2018nonverbal, altmann2020associations}. Greater synchrony strength was associated with lower post-session negative affect. In addition, participants whose synchronous head movements occurred more strongly before the synchronous head movements of the listeners reported lower post-session negative affect and state anxiety.

Participants’ perceptions of the simulated robot were also associated with the interaction dynamics of the sessions. Sessions in which the simulated robot was perceived as autonomous exhibited more instances of genuine synchrony and stronger peak synchrony. More negative attitudes toward robots were likewise associated with a greater proportion of genuine synchrony and stronger peak synchrony.

Together, these findings provide evidence that dyadic synchrony in simulated robot-facilitated anxiety support groups exhibits relationships with therapeutic outcomes similar to those reported in human-facilitated psychotherapy. The findings also demonstrate that participants’ preexisting attitudes toward robots and beliefs about the simulated robot facilitator shape their nonverbal behavior. Future research on robot- and AI-mediated support groups should consider both the agent’s therapeutic capabilities and participants’ perceptions of and responses to the agent technology.

Our work makes three primary contributions towards the development of robot-facilitated support groups.

\begin{enumerate}
    \item We show that robot-facilitated groups exhibit interaction dynamics similar to those of human-facilitated groups, with comparable associations with therapeutic outcomes.
    \item We demonstrate the importance of supporting group processes rather than focusing only on content delivery.
    \item We connect synchrony, therapeutic outcomes, perceived autonomy, and attitudes toward robots in a way that can guide the design and evaluation of future robot- and AI-mediated group interventions, including the development of interaction-sensitive objective functions for autonomous facilitators.
\end{enumerate}

\section{Related Work}
\subsection{Synchrony in Therapeutic Contexts}

Interpersonal synchrony is widely defined as spontaneous temporal coordination between people in behavior or physiology \cite{schmidt2012measuring, palumbo2017interpersonal}. This behavior can be simultaneous, but is often delayed in response. In both dyadic and individual contexts, interpersonal synchrony has been associated with a range of interpersonal and intrapersonal outcomes. For example, stranger dyads who engaged in self-disclosure exhibited greater behavioral synchrony than those who did not \cite{vacharkulksemsuk2012strangers}. Another experiment using Motion Energy Analysis (MEA), a simple measure of movement in a video feed, found that experimentally increased synchrony led to greater positive affect and reduced negative affect among participants \cite{tschacher2014nonverbal}.

A growing body of research has examined the relationship between interpersonal synchrony and both dyadic and individual outcomes in therapeutic settings. Research using MEA has generally found that greater synchrony is associated with greater symptom reduction \cite{ramseyer2011nonverbal, paulick2018nonverbal, altmann2020associations}. Synchrony has also been linked to therapeutic alliance and therapy dropout \cite{schoenherr2021influence, paulick2018nonverbal}. However, associations between synchrony and therapeutic outcomes vary across modalities. Greater MEA-based synchrony has generally been associated with more positive therapeutic outcomes \cite{ramseyer2011nonverbal, paulick2018nonverbal, altmann2020associations}, whereas greater vocal synchrony has been associated with more negative therapeutic outcomes \cite{schoenherr2021influence}. Head-movement synchrony, meanwhile, appears to be associated with multi-session therapist-patient relationship outcomes rather than single-session therapist-patient relationship outcomes, in contrast to broader measures of body synchrony \cite{ramseyer2014nonverbal}. The directionality of synchrony also matters: patient-led synchrony has generally been associated with poorer therapeutic outcomes \cite{altmann2020associations, schoenherr2021influence}.

There is some work examining multiparty synchrony outside these settings \cite{ellamil2016influences}. Multiple work have found both group variance and dyadic synchrony within groups to be strong influences on movement synchrony. For example, \citet{dale2020body} looked at body synchrony in triadic open conversations, measured by taking the windowed cross-correlation peaks with surrogate generation, as in our work, between the two dyads in each triad. They then examined both the dyads themselves and the covarying synchrony measures between dyads. The authors found that both triads and their component pairs showed small but significant body synchrony, albeit not synchrony that was significantly correlated to outcomes (perceptions of group members, demographics, experimental condition, etc.). Another study analyzing multiparty synchrony had groups of individuals in rocking chairs rock in two conditions: one with their eyes closed and one with their eyes open. Using a clustered phase model, they found that group-level synchrony was modestly increased in the eyes-open condition compared with the eyes-closed condition. Although they report no experimental results for dyadic synchrony, they posit that dyadic synchrony measures within groups are important controls, as they represent different psychological processes from group cohesion \cite{dale2020body}. Together, these works indicate that investigating both dyadic and group variance in synchrony is important to understanding group-level outcomes in non-therapeutic or support group contexts. We expand on these methods by performing dyadic comparisons with random group-level effects.

Few analyses have examined therapeutic synchrony beyond dyadic therapist-patient interactions. One exception is \citet{gordon2020physiological}, which examined group drumming sessions and found that interbeat interval and behavioral synchrony predicted greater perceived group cohesion. In a therapeutic context, \citet{prakash2025two} examined movement synchrony during group dance/movement therapy but did not investigate its relationship with therapeutic outcomes. Thus, despite growing evidence for the importance of synchrony in dyadic psychotherapy, relatively little is known about interpersonal synchrony and its relationship to outcomes in group psychotherapy. This gap motivates our investigation of synchrony within group therapy.

\subsection{Facilitation of Psychotherapy by Agents}

The use of agents as psychotherapy facilitators has received considerable research attention. Virtual and robotic agents offer a potentially scalable form of support in an area of healthcare where demand frequently exceeds the availability of trained providers. Importantly, these systems need not replace human therapists to be useful. Instead, embodied agents occupy a range of supportive roles, such as guiding structured practice exercises, encouraging disclosures and engagement, and facilitating interaction among participants. Their ability to deliver consistent prompts and behaviors also makes them useful for interventions that require repeated or highly structured interactions.

Most previous work has focused on dyadic therapeutic interactions between a single participant and a virtual agent \cite{li2025virtual, kang2010virtual, woo2024adaptive, burton2016pilot, frischholz2024examining, shidara2024adapting, devault2014simsensei, kian2025using}. In these settings, the agent primarily manages its interaction with one participant, allowing researchers to focus on factors such as rapport, disclosure, responsiveness, and therapeutic engagement. Several investigations, however, have examined embodied agents as facilitators in group therapy sessions \cite{kian2026embodied, hyekang2025multimodal, kase2019tele, islam2025comparing}. Group settings introduce additional challenges because the agent must respond to individual participants as well as to interpersonal and group-level dynamics. Facilitation requires managing turn-taking, encouraging balanced participation, maintaining conversational structure, and responding appropriately to differences in participants' engagement or affect \cite{bostrom1993group}.

Across both dyadic and group settings, therapeutic agents have been explored with a wide range of populations and clinical needs. For children with Autism Spectrum Disorder, for example, embodied agents have been explored as tools for supporting social interaction and cognitive interventions \cite{shi2022toward, scassellati2018improving, scassellati2012robots}. Agents have also been studied in interventions involving older adults and individuals and those with dementia \cite{bemelmans2015effectiveness, tapus2009use, lima2025promoting}, as well as individuals experiencing anxiety and stress \cite{kian2026embodied, hyekang2025multimodal, islam2025comparing}, depression \cite{burton2016pilot}, and substance use disorder \cite{li2025virtual}. The breadth of these applications suggests that embodied agents support therapeutic interactions across substantially different user populations and treatment contexts. Our work builds on the existing literature by examining how the agent's role affects intervention quality.

\section{Methods}
\subsection{Data}

The study data were collected as part of a Wizard-of-Oz (WoZ) experiment examining participants’ perceptions of a simulated robot facilitator of a virtual mental health support group [redacted]. The facilitator guided groups of university students through discussions about stressors associated with academic life. The study was conducted in English in the United States.

\subsubsection{Simulated Robot Design}
The simulated robot’s facial expressions were designed using Facial Action Units (FAUs) \cite{ekman1978facial}  by varying the shape and position of the eyes, eyebrows, and mouth. These facial components were combined with skeletal animations to produce backchannel behaviors: chuckling, head nodding, showing interest, appearing pensive, appearing questioning, appearing sad, shaking its head, sighing, nodding slowly, and appearing surprised. The robot displayed one animation state at a time. Idle breathing served as the default state, whereas speaking and backchanneling were actively selected by the WoZ operator. While participants were speaking, the WoZ operator selected backchanneling responses that were appropriate to the content, such as chuckling in response to humorous remarks or nodding to convey engagement with the speaker. Unity’s Animator Controller, a finite-state machine, managed the robot's animation states and blended animations between state transitions, enabling smooth transitions among the idle breathing, speech, and backchanneling behaviors. Throughout the interaction, the robot’s head remained oriented straight ahead toward the simulated camera, except when it was performing backchanneling behaviors such as head shaking.

\subsubsection{Session Structure}
\begin{figure}[H]
    \centering
    \includegraphics[width=0.5\linewidth]{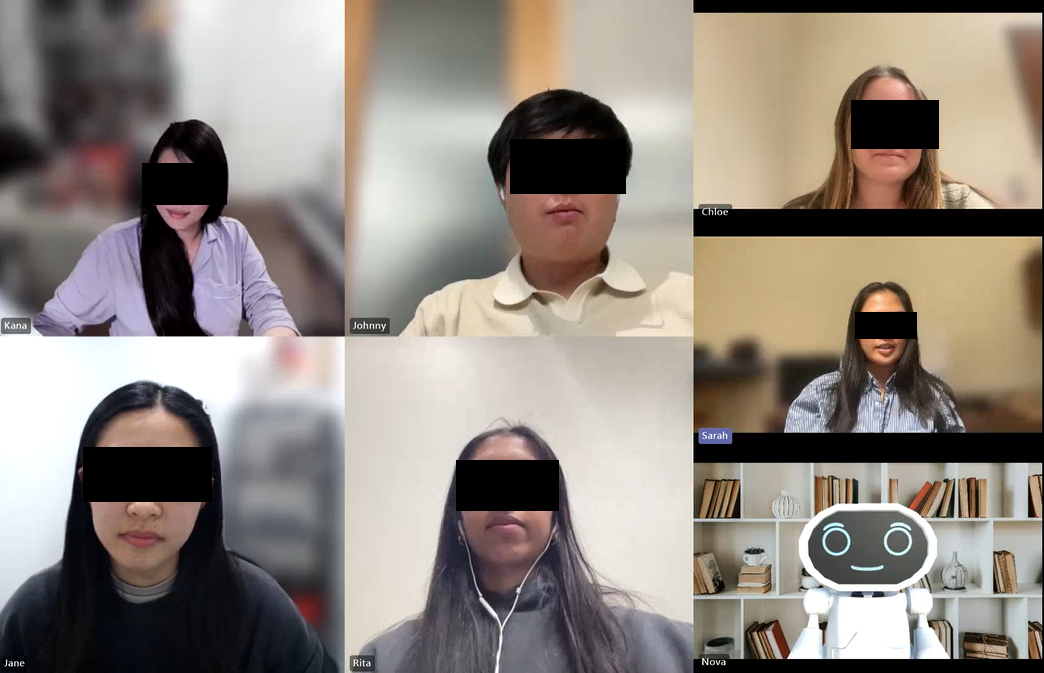}
    \caption{A snapshot of a mock robot-facilitated support group session on Microsoft Teams. Displayed names are fictional.}
    \label{fig:meeting}
    \Description{A screenshot of a Microsoft Teams meeting with one robot and 6 human participants.}
\end{figure}

Support group sessions were held remotely via Microsoft Teams, as shown in Figure \ref{fig:meeting}. A researcher initiated each session, then turned their camera and microphone off once the robot joined and began facilitating the anxiety support group. Each session comprised four discussion stages that progressed from relatively simple prompts to increasingly personal or in-depth topics. At each stage, the robot facilitator introduced the topic and invited participants to contribute. Students followed a discussion format with each participant speaking at least once in response to every discussion prompt. 

Audio and video data were recorded throughout the session, and participants completed self-report questionnaires before and after the session.

Because the simulated robot facilitator was operated using a WoZ protocol, participants, in the autonomous frame condition, were debriefed about the deception at the conclusion of the study. The study was approved by the [Redacted] Institutional Review Board (IRB [Redacted]), and participants received US\$50 in compensation.

\subsubsection{Instruments}

Surveys were administered to all participants after each session. The analysis used the following instruments:
\\
\textit{Positive and Negative Affect Schedule Expanded Form (PANAS-X).} The 60-item PANAS \cite{watson1994panas} assesses participants’ positive and negative affect. The General Dimension Scales consist of a 10-item Positive Affect (PA) and a 10-item Negative Affect (NA) scale. Items are rated on a five-point Likert scale ranging from 1 (\textit{very slightly or not at all}) to 5 (\textit{extremely}) and are used to calculate separate positive- and negative-affect scores. The instrument was administered both pre- and post-session.

\textit{State-Trait Anxiety Inventory (STAI).} The 40-item STAI \cite{spielberger1971state} assesses participants’ tendency for anxiety in addition to their current anxiety. Items are rated on a four-point Likert scale ranging from 1 (\textit{not at all}) to 4 (\textit{very much so}) for state anxiety, and 1 (\textit{almost never}) to 4 (\textit{almost always}) for trait anxiety. The scale is split into a 20-item state anxiety and a 20-item trait anxiety measure. The trait was measured pre-session, and only the state was administered pre- to post-session.

\textit{Negative Attitudes toward Robots Scale (NARS).} The 14-item NARS \cite{nomura2006measurement} assesses participants’ attitudes toward robots. Items are rated on a five-point Likert scale ranging from 1 (\textit{strongly disagree}) to 5 (\textit{strongly agree}). This instrument was administered pre-session.

\subsubsection{Participant Demographics}

Eligible participants were between 18 and 39 years old, proficient in English, and had normal or corrected-to-normal hearing and vision. Individuals were excluded if they scored 11 or higher on the Patient Health Questionnaire (PHQ-8) \cite{kroenke2009phq} or had a Word Error Rate above 8\% on the automatic speech-recognition assessment.
Following screening, the final sample comprised 96 participants recruited from a university in a large U.S. city. The 96 participants were distributed across 18 anxiety support group sessions, with an average of 5.33 participants per session. Participants’ mean age was 32.39 years ($SD=3.22$). 51\% of participants identified as men, 42\% as women, and 3\% as non-binary; gender data for the remaining 4\% were unreported. Participants primarily identified as Asian (80.2\%), followed by Latine (5.2\%), White (4.2\%), Black or African American (3.1\%), multiracial (3.1\%), and North African (2.1\%); race or ethnicity data for the remaining participants were unreported (5.2\%). The highest education levels reported were a bachelor's degree (38.5\%), a master's degree (35.4\%), a high school diploma or General Education Development (GED) credential (11.5\%), some college education (11.5\%), and a doctoral degree for the remaining participants (3.1\%).

\begin{figure}[]
    \centering

    \begin{subfigure}[b]{0.48\textwidth}
        \centering
        \includegraphics[width=\linewidth]{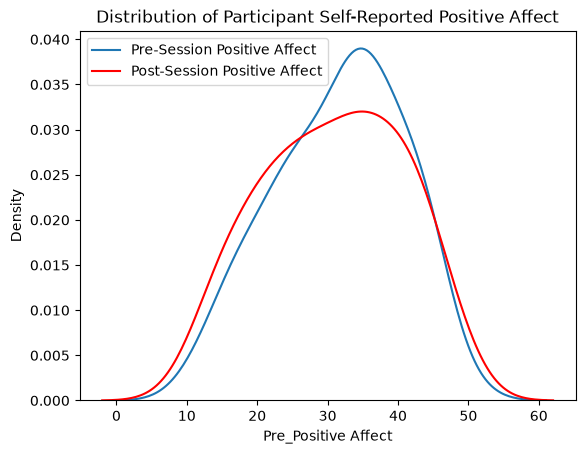}
        \caption{Positive Affect}
        \Description{Density plot of pre- and post-session positive affect. Scores are normally distributed around the center.}
        \label{fig:positive_affect}
    \end{subfigure}
    \hfill
    \begin{subfigure}[b]{0.48\textwidth}
        \centering
        \includegraphics[width=\linewidth]{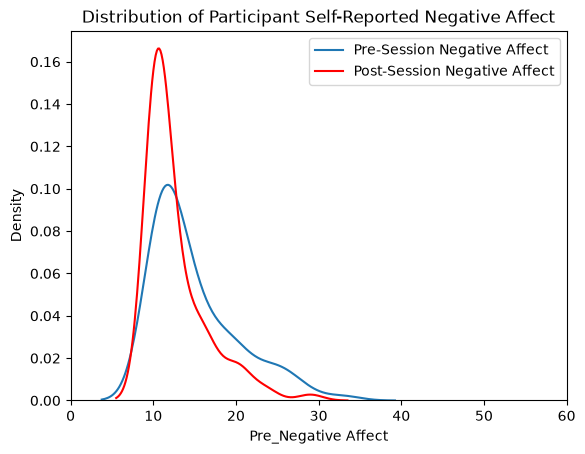}
        \caption{Negative Affect}
        \Description{Density plot of pre- and post-session negative affect. Scores are skewed towards low amounts of negative affect.}
        \label{fig:negative_affect}
    \end{subfigure}

    \medskip

    \begin{subfigure}[b]{0.48\textwidth}
        \centering
        \includegraphics[width=\linewidth]{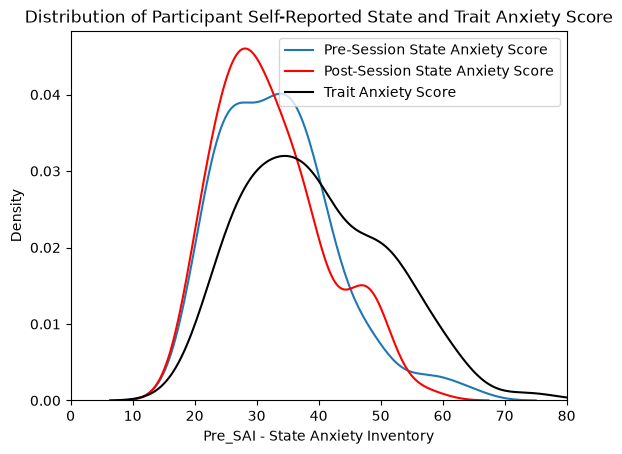}
        \caption{State and Trait Anxiety}
        \Description{Density plot of pre- and post-session state anxiety and trait anxiety. All scores are normally distributed around the center.}
        \label{fig:anxiety}
    \end{subfigure}
    \hfill
    \begin{subfigure}[b]{0.48\textwidth}
        \centering
        \includegraphics[width=\linewidth]{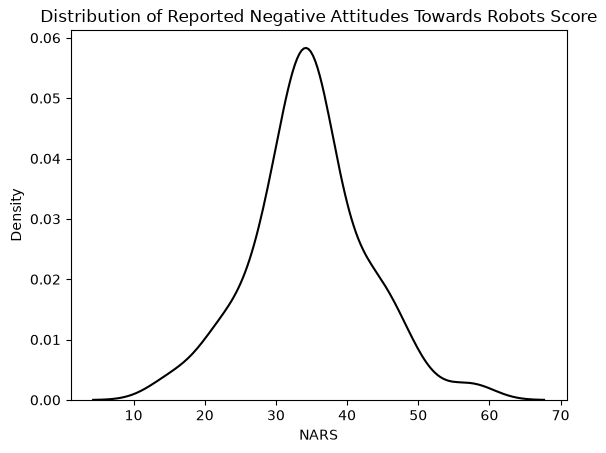}
        \caption{Negative Attitudes Towards Robots}
        \Description{Density plot of NARS. Scores are normally distributed around the center.}
        \label{fig:nars}
    \end{subfigure}

    \caption{Distribution of scores from reported affect, state and trait anxiety, and NARS questionnaires.}
    \label{fig:scores}
\end{figure}

 Participants generally had some knowledge of robots, with 43.7\% reporting having a little, 40.6\% having some, and the remainder having a lot. Participants generally had little to no experience with support groups, with 69.8\% having attended 0 support group sessions, 28.1\% having attended 1 to 4 sessions, and 2.1\% having attended 5 or more sessions. The distribution of participants' positive and negative affect is shown via density plots in Figures \ref{fig:positive_affect} and \ref{fig:negative_affect}, respectively, for both pre- and post-session measures. Both state and trait anxiety measures were collected, as shown in Figure \ref{fig:anxiety}. Finally, the participants' reported negative attitudes toward robots are shown in Figure \ref{fig:nars}. 

\subsection{Synchronous Head Movement}

Low-level synchrony in head movements was defined using head pose defined by pitch, roll, and yaw Euler angles, tracked by OpenFace \cite{baltruvsaitis20182}. For each study session participant dyad and each head-rotation variable, we took the signal of the speaker's utterance as $x = \{x_1, x_2, ..., x_n\}$ and a listener's corresponding signal as $y=\{y_1, y_2, ..., y_n\}$. We first excluded frames for which OpenFace's pose-estimation confidence was $<0.70$. This threshold was chosen empirically to balance data quality and the number of frames included. To reduce high-frequency measurement noise, the remaining observations were aggregated into non-overlapping 250-ms bins, or about 8 frames, by averaging the valid observations within each bin. This window size was selected because, as in \citet{shen2012effects}, micro expressions are detectable by 200-ms and stabilize by 300-ms, motivating our choice of a point between the two.

We then applied a rolling 3-second window to the resulting time series. We chose the 3-second duration because it is consistent with prior work using windowed cross-correlation to quantify dyadic head-movement synchrony; such analyses have selected 3 seconds to balance the expected temporal duration of interpersonal coordination with the need to capture changes in synchrony over the course of an interaction \cite{richardson2007art}. Within each window, the signals were mean-centered.

We then calculated the linear cross-correlation between the two signals using the Fast Fourier Transform. For lag $k \in\{-n-1, \ldots, n-1\}$, the unnormalized cross-correlation was defined as

\begin{align}
    C(k)=\sum_{i\in\{I_k\}}x'_{i}y_{i-k}
\end{align}

where $I_k$ is the set of indices for which both $x'_i$ and $y'_{i-k}$ are observed. The resulting correlations were normalized using the corresponding standard deviations, where $r(k)$ corresponds to the Pearson correlation at each lag.

For each 3-second window, we identified the peak correlation and its corresponding lag as

\begin{align}
    r_{\max} &= \max_k r(k) &
    \mathrm{lag}_{\max} &= \arg\max_k r(k)
\end{align}

\begin{figure}[H]
    \centering
    \includegraphics[width=0.5\linewidth]{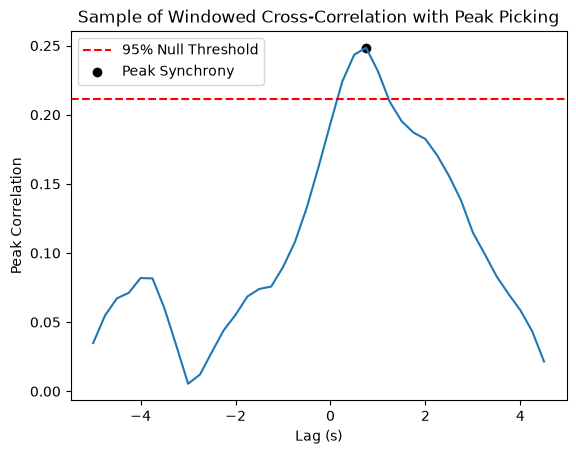}
    \caption{For an example dyad in the dataset, the cross correlation between the dyad's head rotation is shown over various lags, with the peak synchrony indicated by a dot and the null threshold in red.}
    \Description{A graph with lag on the x axis, peak correlation on the y axis, and a line with each lag's corresponding correlation. There is a threshold for pseudo-synchrony, and a peak indicated that is above that threshold.}    
    \label{fig:wcc_peak_picking}
\end{figure}

To distinguish genuine synchrony from pseudosynchrony (the instances in which the peak correlation did not surpass the 95\% surrogate threshold), we generated 100 surrogate dyads by independently randomizing the temporal ordering of the two signals and repeating the same cross-correlation procedure for each surrogate. This produced a null distribution of surrogate peak correlations, $r^{\underset{(s)}{max}}$ against which the observed peak was compared. A peak was classified as synchronous when $r_{max}$ excluded the 95th percentile of the surrogate peak-correlation distribution. This process is visualized in Figure \ref{fig:wcc_peak_picking}.

\section{Results}
We first investigated \textbf{RQ1}; if there are relationships between head-movement synchrony and therapeutic outcomes in simulated robot-facilitated support groups. To address this question, we examine head-movement synchrony using peak correlation values, the lag between speaker and listener at the peak synchrony, and the percentage of the interaction that is not pseudosynchrony (defined as $>95\%$ of the peak correlation of the surrogate distribution). The outcome variables include positive and negative affect, as measured by the PANAS questionnaire, and state anxiety, as measured by the state portion of the STAI questionnaire.  

We then sought to answer \textbf{RQ2}; if participants' perceptions of the simulated robot are associated with differences in head-movement synchrony. We again use peak synchrony strength, synchrony lag, and percent non-psuedosynchrony as measures of synchrony. To assess participants' perceptions of the simulated robot facilitator, we use the robot condition (perceived as autonomous or human-operated) and the NARS questionnaire. 

\subsection{Relationship Between Synchrony and Therapeutic Outcomes}
To investigate our first research question, whether the relationship between head-movement synchrony and outcomes in robot-facilitated contexts resembles those in human-facilitated contexts, we examined the relationship between synchrony and post-session outcomes. Before assessing this relationship, however, we first examined the associations between pre-session and post-session measures of positive affect, negative affect, and state anxiety.

\begin{figure}[H]
    \centering
    \includegraphics[width=0.8\linewidth]{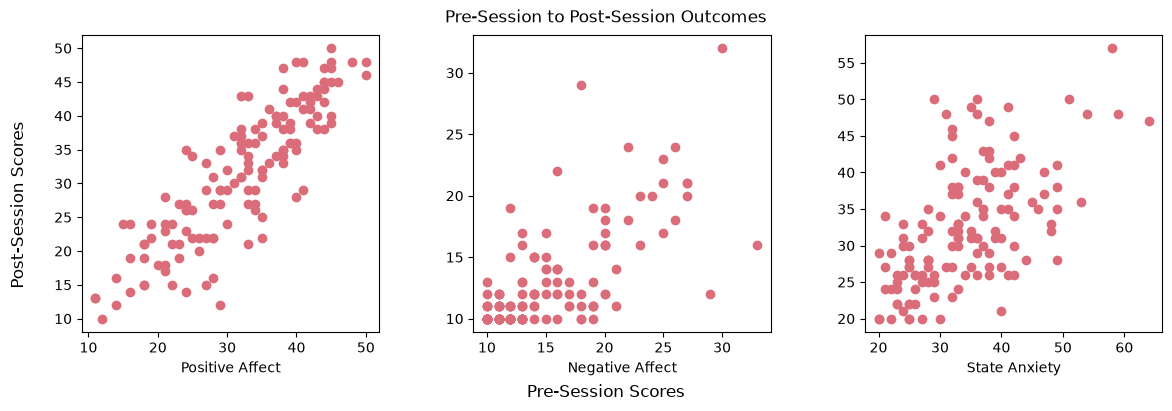}
    \caption{Pre-Session vs Post-Session Outcomes}
    \Description{Three scatter plots side-by-side, all with pre-session on the x-axis and post-session on the y-axis. Pre-session positive affect, negative affect, and state anxiety are all positively and visually correlated with their respective post-session scores. Positive affect is the strongest linear correlation of the three.}
    \label{fig:pre-post-outcomes}
\end{figure}

We found strong correlations between all corresponding pre- and post-session measures. Pre-session positive affect was positively correlated with post-session positive affect, $r(96) = 0.841$, $p < 0.001$. Pre-session negative affect was positively correlated with post-session negative affect, $r(96) = 0.682$, $p < 0.001$. Finally, pre-session state anxiety was positively correlated with post-session state anxiety, $r(96) = 0.611$, $p < 0.001$. These results are visualized in Figure \ref{fig:pre-post-outcomes}.

To estimate the associations between synchrony and post-session outcomes while accounting for baseline differences, we included the corresponding pre-session measure as a covariate in each model. Synchrony variables were averaged within speaker-dyad pairs because the outcomes were measured at the post-session level. Thus, aggregating the synchrony predictors at the outcome level aligned them with the outcome level and avoided treating repeated within-session observations as independent. Although this aggregation reduces within-speaker variability in synchrony, our analyses focused on differences in speakers’ average synchrony across sessions rather than moment-to-moment variation. The models included fixed effects for the corresponding pre-session measure and synchrony variable. To explicitly account for the nested structure of the data and the non-independence of observations within the same session, each model included a random intercept for session. Continuous questionnaire and synchrony variables were z-standardized. Consequently, coefficients for continuous predictors represent the expected difference in the outcome, in standard-deviation units, associated with a one-standard-deviation increase in the predictor, conditional on the other terms in the model. Models containing statistically significant synchrony coefficients are summarized in Table \ref{tab:lmme_sync}; complete results, including models with nonsignificant synchrony coefficients, are reported in Appendix \ref{app:A}.

\begin{table}[H]
\centering
\resizebox{\columnwidth}{!}{%
\begin{tabular}{|l|l|l|ll|}
\hline
\textbf{Outcome Variable}                      & \textbf{Model Term}     & \textbf{Estimate} & \multicolumn{1}{l|}{\textbf{p}} & \textbf{95\% CI} \\ \hline
                                               & Pre-Session Negative Affect       & 0.673                & \multicolumn{1}{l|}{< 0.001}    & [0.528, 0.818]    \\ \cline{2-5} 
                                               & Average Synchronous Peak Strength & -0.168               & \multicolumn{1}{l|}{0.023}      & [-0.313, -0.023]  \\ \cline{2-5} 
\multirow{-3}{*}{Post-Session Negative Affect} & Session-Level Random-Intercept Variance      & < 0.001              & \multicolumn{2}{l|}{\cellcolor[HTML]{B7B7B7}}       \\ \hline
                                               & Pre-Session Negative Affect       & 0.635                & \multicolumn{1}{l|}{< 0.001}    & [0.502, 0.767]    \\ \cline{2-5} 
                                               & Average Synchronous Lag           & -0.345               & \multicolumn{1}{l|}{< 0.001}    & [-0.478, -0.213]  \\ \cline{2-5} 
\multirow{-3}{*}{Post-Session Negative Affect} & Session-Level Random-Intercept Variance      & < 0.001              & \multicolumn{2}{l|}{\cellcolor[HTML]{B7B7B7}}       \\ \hline
                                               & Pre-State Anxiety                 & 0.598                & \multicolumn{1}{l|}{< 0.001}    & [0.454, 0.742]    \\ \cline{2-5} 
                                               & Average Synchronous Lag           & -0.249               & \multicolumn{1}{l|}{0.001}      & [-0.393, -0.104]  \\ \cline{2-5} 
\multirow{-3}{*}{Post-State Anxiety}           & Session-Level Random-Intercept Variance      & 0.11                 & \multicolumn{2}{l|}{\cellcolor[HTML]{B7B7B7}}       \\ \hline
\end{tabular}%
}
\caption{Linear mixed-effects models showing statistically significant associations (p-value of $\beta$ $<0.05$) between movement synchrony and post-session outcome scores.}
\label{tab:lmme_sync}
\end{table}

After fitting the LMMs, we conducted a sensitivity analysis to assess whether the study was sufficiently powered to detect the significant effects. The associations of synchrony lag with negative affect and state anxiety met the 80\% power threshold, with minimum detectable effects (MDEs) of 0.189 and 0.207, respectively. In contrast, the association between synchrony strength and negative affect did not meet this threshold (MDE = 0.207). This finding should therefore be interpreted with caution and investigated further in a study with a larger sample size and greater statistical power.

\subsection{Impact of the Participants' Perceptions of the Robot on Synchrony}

To address our second research question, whether participants' perceptions of the simulated robot are associated with differences in head-movement synchrony, we first examined whether robot condition and NARS were associated with session outcomes. We then examined the relationships among these predictors, movement synchrony, and session outcomes.

In the initial outcome models, robot condition was significantly associated with post-session state anxiety, but not with post-session negative or positive affect. NARS was not significantly associated with any of the three post-session outcomes. To facilitate comparisons across measures, all outcome variables and NARS scores were standardized. Separate LMMs were fitted for each post-session outcome, with either robot condition or NARS as the main predictor, the corresponding pre-session score as a covariate, and participant-level random effects. Participant-level random intercepts accounted for repeated outcome observations within participants, whereas session-level random intercepts accounted for the clustering of speaker-level synchrony observations within sessions.

Robot condition was not significantly associated with post-session negative or positive affect. For post-session negative affect, the contrast between the perceived human-operated condition and the perceived autonomous condition was nonsignificant ($\beta = 0.181$, $SE = 0.156$, $p = 0.246$). Pre-session negative affect was positively associated with post-session negative affect ($\beta = 0.680$, $SE = 0.006$, $p < 0.001$). The participant-level intraclass correlation coefficient was 0.27.

Similarly, for post-session positive affect, the contrast between the perceived human-operated condition and the perceived autonomous condition was nonsignificant ($\beta = 0.071$, $SE = 0.116$, $p = 0.543$). Pre-session positive affect was positively associated with post-session positive affect ($\beta = 0.837$, $SE = 0.005$, $p < 0.001$). The participant-level intraclass correlation coefficient was 0.15.

In contrast, robot condition was significantly associated with post-session state anxiety. Relative to participants in the perceived autonomous robot condition, participants in the perceived human-operated condition had post-session state anxiety scores that were 0.217 standard deviations higher after adjustment for pre-session state anxiety ($\beta = 0.217$, $SE = 0.066$, $p = 0.001$). Sensitivity analysis indicates sufficient power for this finding at MDE 80\% at 0.185. Pre-session state anxiety was also positively associated with post-session state anxiety ($\beta = 0.611$, $SE = 0.081$, $p < 0.001$). The participant-level intraclass correlation coefficient was 0.31.

NARS was not significantly associated with post-session negative affect ($\beta = -0.088$, $SE = 0.074$, $p = 0.236$), positive affect ($\beta = -0.035$, $SE = 0.057$, $p = 0.542$), or state anxiety ($\beta = 0.033$, $SE = 0.037$, $p = 0.375$). In all three models, the corresponding pre-session score significantly associated with the post-session score: negative affect ($\beta = 0.694$, $SE = 0.011$, $p < 0.001$), positive affect ($\beta = 0.834$, SE = 0.009, $p < 0.001$), and state anxiety ($\beta = 0.608$, $SE = 0.082$, $p < 0.001$). The participant-level intraclass correlation coefficients were 0.27, 0.15, and 0.32 for negative affect, positive affect, and state anxiety, respectively.

Having found that robot condition was associated with post-session state anxiety, but not with post-session affect, and that NARS was not significantly associated with any post-session outcome, we next examined whether robot condition and NARS were associated with movement synchrony. Table \ref{tab:lmme_robot} presents the results of LMMs predicting synchrony from robot condition and NARS, with a random intercept for session. Complete results are reported in Appendix \ref{app:B}.

\begin{table}[H]
\centering
\resizebox{\columnwidth}{!}{%
\begin{tabular}{|l|l|l|ll|}
\hline
\textbf{Outcome Variable}                              & \textbf{Coefficient Variable} & \textbf{Coefficient ($\beta$)} & \multicolumn{1}{l|}{\textbf{p}} & \textbf{95\% CI} \\ \hline
 & Intercept (Perceived Autonomous Condition) & 0.372  & \multicolumn{1}{l|}{0.060} & [-0.016, 0.760]  \\ \cline{2-5} 
 & Perceived Human vs.\ Percieved Autonomous Condition                  & -0.577 & \multicolumn{1}{l|}{0.021} & [-1.067, -0.086] \\ \cline{2-5} 
 & NARS                                       & 0.191  & \multicolumn{1}{l|}{0.042} & [0.007, 0.375]   \\ \cline{2-5} 
\multirow{-4}{*}{Average Synchronous Peak Strength}    & Session-Level Random-Intercept Variance  & 0.111                & \multicolumn{2}{l|}{\cellcolor[HTML]{B7B7B7}}       \\ \hline
 & Perceived Autonomous Condition (intercept) & 0.333  & \multicolumn{1}{l|}{0.096} & [-0.059, 0.725]  \\ \cline{2-5} 
 & Perceived Human-Controlled Condition                  & -0.520 & \multicolumn{1}{l|}{0.040} & [-1.016, -0.024] \\ \cline{2-5} 
 & NARS                                       & 0.205  & \multicolumn{1}{l|}{0.031} & [0.019, 0.391]   \\ \cline{2-5} 
\multirow{-4}{*}{\% of Interaction That is Synchronous} & Session-Level Random-Intercept Variance  & 0.115                & \multicolumn{2}{l|}{\cellcolor[HTML]{B7B7B7}}       \\ \hline
\end{tabular}%
}
\caption{Linear mixed-effects models showing statistically significant associations (p-value of $\beta$ $<0.05$) of perceived robot condition and NARS scores with movement synchrony.}
\label{tab:lmme_robot}
\end{table}

A post hoc sensitivity analysis indicated that the study had 80\% power to detect the association between the robot-perceived autonomy condition and state anxiety, with an MDE of 0.185. None of the other associations met the 80\% power threshold, likely because the analysis included only 18 sessions. Although the findings were promising and statistically significant, they should be interpreted cautiously and replicated in a larger sample.

\section{Discussion}
\subsection{Dyadic Synchrony and Therapeutic Outcomes}

Addressing our first research question on the relationship between head-movement synchrony and outcomes, our findings indicate that head-movement synchrony predicts both post-session negative affect and post-session state anxiety. Through linear mixed-effects modeling, we found that, controlling for pre-session negative affect, stronger peak synchrony was correlated with lower post-session negative affect. We also found that, controlling for pre-session measures, more following behavior, in which the speaker responds to the listener through their movements, was correlated with greater negative affect and state anxiety after the group session. These findings generally support the view that dyadic synchrony in group contexts is relevant to therapeutic outcomes.

Our findings both support and expand upon the existing literature. \citet{ramseyer2011nonverbal} found that head-movement synchrony between therapists and patients was associated with macro-level outcomes. Although we analyzed only single sessions, our findings suggest that, in group settings, head-movement synchrony between participants is also associated with micro-level outcomes such as state anxiety and affect. \citet{altmann2020associations} and \citet{schoenherr2021influence} both concluded that patient-led synchrony was related to poorer outcomes. Although we did not consider synchrony between the robot facilitator and participants, we found that speaking participants experienced more negative outcomes when responding to the movements of listening participants, the inverse of the patient-led pattern in a patient-therapist relationship. These findings reinforce the association between synchrony and therapeutic outcomes and extend it to synchrony among group participants.

The relevance of our findings is twofold: they show that robot-facilitated support groups align with established trends in the relationship between synchrony and outcomes, and they demonstrate that synchrony remains an important measure of interaction quality beyond dyadic therapist-patient interactions. These findings call for further investigation into how robot facilitators encourage beneficial forms of participant synchrony and whether similar patterns occur in other group-therapy contexts, including those with human facilitators.

\subsection{Perceptions of Robots and Dyadic Movement}

Although we found no significant associations between robot condition or NARS with the post-session affect or anxiety outcomes, the perceived autonomous robot condition was significantly associated with post-session state anxiety, but not with positive or negative affect. Participants in the perceived human-operated condition reported higher post-session state anxiety than those in the perceived autonomous condition after controlling for pre-session state anxiety. 

We also identified several relationships among the robot condition, NARS scores, and participant synchrony. Both the percentage of the interaction classified as genuine synchrony and the strength of synchronous peaks were higher in the perceived autonomous robot condition than in the perceived human-operated condition. Additionally, more negative attitudes toward robots were associated with a higher percentage of genuine synchrony and stronger synchronous peaks.

These results indicate that perceptions of facilitator autonomy were associated with both participant synchrony and post-session state anxiety. In human-facilitated groups, the identity of the facilitator has been found to significantly impact head movement, indicating the importance of facilitator identity matched previous work \cite{chew2024human}. One possible explanation is that participants focused more on one another because they perceived the robot as less human or less competent as a facilitator. However, our analyses do not establish whether differences in synchrony explain the association between perceived robot condition and state anxiety. These findings support the need for further investigation into the use of autonomous robots in robot-facilitated support groups.

We also found that NARS scores were positively associated with synchrony, despite not being significantly associated with post-session outcomes. One possible explanation is that participants with more negative attitudes toward robots paid less attention to the robot and more attention to other group members. Prior research characterizes group-session facilitators primarily as facilitators of structure rather than participants \cite{bostrom1993group}, which may explain the observed pattern.

These findings indicate that future work involving robot-facilitated support groups must consider not only robot and interaction design, but also how the robot is framed and perceived. In particular, further research is needed to determine why perceiving the facilitator as human-operated was associated with higher post-session state anxiety despite being associated with lower participant synchrony.

\subsection{Limitations}

This work has several limitations. The support group facilitators who controlled the WoZ simulated robot interface, although extensively trained, were not licensed facilitators. This characteristic limits the experiment's relevance to psychotherapy, although the findings remain applicable to support group contexts. The experiment also involved only single sessions, limiting our ability to assess long-term effects and potentially introducing novelty bias. The single cultural context further limits the generalizability of the results.

The analysis we performed also has some limitations. Head rotation was estimated using OpenFace. Although we filtered the data to retain only high-quality observations, model accuracy may still be affected by video quality and other factors. The quality of participants' Internet connections potentially affected synchrony latency and influenced results related to synchrony lag. Additionally, there were only 18 sessions with 96 participants, with sensitivity analysis showing occasional underpowered statistical significance. It is important that these results are explored in a larger number of groups.

\subsection{Future Work}

These findings motivate further research on both synchrony in support groups and robot-facilitated support groups. Little existing research examines synchrony among participants in support groups. Although our analysis focused on a (simulated) robot-facilitated context, future work can examine whether similar patterns occur in groups with human facilitators. Direct comparisons among autonomous robots, teleoperated robots, and human facilitators help to distinguish the effects of facilitator embodiment from those of perceived autonomy and facilitation style.

Future research can also investigate the mechanisms underlying the observed associations. In particular, the contrasting relationships between peak synchrony and who was responding two who in synchronous movements suggest that different aspects of synchrony have different implications for participants. Strong shared movement may reflect rapport or engagement, whereas repeatedly responding to another participant rather than the listener responding to the speaker may reflect withdrawal, uncertainty, or an unequal interaction dynamic. Combining movement measures with turn-taking patterns, speech content, self-reported engagement, and participants' perceptions of other group members helps clarify these possibilities. Examining additional nonverbal behaviors, such as facial expressions, posture, gestures, and vocal features, also provides a more complete account of interpersonal coordination.

The findings also reveal positive associations between synchrony and negative attitudes toward robots. Future work can examine how different robot embodiments, behaviors, and introductions affect perceptions of the robot and, consequently, synchrony between participants. Research involving repeated sessions with the same participants offers an opportunity to reduce novelty effects and determine whether robot perceptions and interaction patterns change over time. Such longitudinal work also supports assessing whether synchrony predicts longer-term outcomes, including continued participation, group cohesion, perceived social support, and changes in psychological well-being.

Finally, experimental designs offer a means to test whether facilitator behaviors encourage forms of synchrony associated with beneficial outcomes. For example, a robot facilitator can adjust prompts, speaking order, or opportunities for reflection in response to group interaction patterns. Because greater synchrony was not uniformly associated with better outcomes in this analysis, treating synchrony as an inherently desirable behavior to maximize may not be warranted. Further research is needed to identify when synchrony reflects healthy engagement and when it may signal potentially unhelpful interpersonal dynamics; such distinctions should inform how synchrony is used to guide adaptive facilitation.

\section{Conclusion}
We examined dyadic synchrony in robot-facilitated support groups for anxiety through a Wizard-of-Oz experimental design. Dyadic synchrony among group participants was calculated from head-rotation data using windowed cross-correlation, peak picking, and surrogate generation. Our findings indicate that head-movement synchrony is associated with post-session negative affect and state anxiety. Stronger peak synchrony was associated with lower negative affect, whereas greater following behavior by speakers was associated with higher negative affect and state anxiety. These contrasting relationships suggest that synchrony is not uniformly beneficial; its implications for therapeutic outcomes depend on the strength, form, and direction of dyadic coordination. These results extend prior research on therapist-patient synchrony by showing that movement coordination among participants is also relevant in a support group context.

Participant synchrony was also associated with perceived autonomy of the robot facilitator and attitudes toward robots. Perceived robot facilitator autonomy was associated with post-session state anxiety, with participants in the perceived human-operated robot condition reporting higher anxiety than those in the perceived autonomous robot condition, whereas attitudes toward robots were not significantly associated with the measured outcomes. Participants in the perceived autonomous condition and those with more negative attitudes toward robots exhibited greater synchrony, highlighting the potential role of robot perception and behavior in shaping group dynamics. Together, these findings support the use of computational measures of synchrony to analyze interactions in robot-facilitated support groups while emphasizing the importance of distinguishing beneficial coordination from potentially unhelpful following behavior. Longitudinal research and comparisons with human-facilitated groups are necessary to determine how robot facilitator design and facilitation strategies can foster productive group interactions and improve participant outcomes over time.

\bibliographystyle{ACM-Reference-Format}
\bibliography{bib}

\appendix

\section{}\label{app:A}

\begin{table}[H]
\centering
\resizebox{\columnwidth}{!}{%
\begin{tabular}{|l|l|l|ll|}
\hline
\textbf{Outcome Variable} &
  \textbf{Coefficient Variable} &
  \textbf{Coefficient} &
  \multicolumn{1}{l|}{\textbf{p}} &
  \textbf{97.5\% CI} \\ \hline
\multicolumn{1}{|c|}{} &
  Pre-Session Negative Affect &
  0.675 &
  \multicolumn{1}{l|}{\textless 0.01} &
  {[}0.529, 0.821{]} \\ \cline{2-5} 
\multicolumn{1}{|c|}{} &
  Average \% of Interaction that is Synchronous &
  -0.141 &
  \multicolumn{1}{l|}{0.052} &
  {[}-0.282, 0.001{]} \\ \cline{2-5} 
\multicolumn{1}{|c|}{\multirow{-3}{*}{Post-Session Negative Affect}} &
  Session-Level Random-Intercept Variance &
  \textless 0.01 &
  \multicolumn{2}{l|}{\cellcolor[HTML]{B7B7B7}} \\ \hline
 &
  Pre-Session Positive Affect &
  0.841 &
  \multicolumn{1}{l|}{\textless 0.01} &
  {[}0.728, 0.954{]} \\ \cline{2-5} 
 &
  Average Synchronous Peak Strength &
  0.003 &
  \multicolumn{1}{l|}{0.953} &
  {[}-0.111, 0.118{]} \\ \cline{2-5} 
\multirow{-3}{*}{Post-Session Positive Affect} &
  Session-Level Random-Intercept Variance &
  \textless 0.01 &
  \multicolumn{2}{l|}{\cellcolor[HTML]{B7B7B7}} \\ \hline
 &
  Pre-Session Positive Affect &
  0.849 &
  \multicolumn{1}{l|}{\textless 0.01} &
  {[}0.737, 0.961{]} \\ \cline{2-5} 
 &
  Average Synchronous Lag &
  0.056 &
  \multicolumn{1}{l|}{0.327} &
  {[}-0.055, 0.167{]} \\ \cline{2-5} 
\multirow{-3}{*}{Post-Session Positive Affect} &
  Session-Level Random-Intercept Variance &
  \textless 0.01 &
  \multicolumn{2}{l|}{\cellcolor[HTML]{B7B7B7}} \\ \hline
 &
  Pre-Session Positive Affect &
  0.835 &
  \multicolumn{1}{l|}{\textless 0.01} &
  {[}0.722, 0.948{]} \\ \cline{2-5} 
 &
  Average \% of Interaction that is Synchronous &
  -0.035 &
  \multicolumn{1}{l|}{0.549} &
  {[}-0.151, 0.080{]} \\ \cline{2-5} 
\multirow{-3}{*}{Post-Session Positive Affect} &
  Session-Level Random-Intercept Variance &
  0.004 &
  \multicolumn{2}{l|}{\cellcolor[HTML]{B7B7B7}} \\ \hline
 &
  Pre-Session Negative Affect &
  0.615 &
  \multicolumn{1}{l|}{\textless 0.01} &
  {[}0.462, 0.768{]} \\ \cline{2-5} 
 &
  Average Synchronous Peak Strength &
  -0.101 &
  \multicolumn{1}{l|}{0.231} &
  {[}-0.267, 0.064{]} \\ \cline{2-5} 
\multirow{-3}{*}{Post-Session State Anxiety} &
  Session-Level Random-Intercept Variance &
  0.075 &
  \multicolumn{2}{l|}{\cellcolor[HTML]{B7B7B7}} \\ \hline
 &
  Pre-Session Negative Affect &
  0.614 &
  \multicolumn{1}{l|}{\textless 0.01} &
  {[}0.460, 0.767{]} \\ \cline{2-5} 
 &
  Average \% of Interaction that is Synchronous &
  -0.038 &
  \multicolumn{1}{l|}{0.659} &
  {[}-0.204, 0.129{]} \\ \cline{2-5} 
\multirow{-3}{*}{Post-Session State Anxiety} &
  Session-Level Random-Intercept Variance &
  0.089 &
  \multicolumn{2}{l|}{\cellcolor[HTML]{B7B7B7}} \\ \hline
\end{tabular}%
}
\caption{Non-Significant Linear mixed-effects models between movement synchrony and post-session outcome scores.}
\label{tab:lmme_sync_nonsig}
\end{table}

\section{}\label{app:B}
\begin{table}[H]
\centering
\resizebox{\columnwidth}{!}{%
\begin{tabular}{|l|l|l|ll|}
\hline
\textbf{Outcome Variable}       & \textbf{Coefficient Variable} & \textbf{Coefficient} & \multicolumn{1}{l|}{\textbf{p}} & \textbf{97.5\% CI} \\ \hline
 & Perceived Autonomous Condition (intercept) & -0.099 & \multicolumn{1}{l|}{0.566} & {[}-0.436, 0.239{]} \\ \cline{2-5} 
 & Perceived Human Condition                  & 0.155  & \multicolumn{1}{l|}{0.470} & {[}-0.266, 0.577{]} \\ \cline{2-5} 
 & NARS                                       & -0.047 & \multicolumn{1}{l|}{0.661} & {[}-0.256, 0.163{]} \\ \cline{2-5} 
\multirow{-4}{*}{Synchrony Lag} & Session-Level Random Effects  & \textless 0.01       & \multicolumn{2}{l|}{\cellcolor[HTML]{B7B7B7}}        \\ \hline
\end{tabular}%
}
\caption{Non-Significant Linear Mixed Effects Model between robot condition, NARS, and synchrony.}
\label{tab:lmme_robot_nonsig}
\end{table}

\end{document}